\documentclass{ws-procs9x6}

\begin{document}

\title{Recent Results from the MAGIC Telescopes}

\author{O. Tibolla$^*$ on behalf of the MAGIC collaboration}

\address{ITPA, Universit\"at W\"urzburg, Campus Hubland Nord,\\
Emil-Fischer-Str. 31 D-97074 W\"urzburg, Germany\\
$^*$E-mail: omar.tibolla@gmail.com ;\\
Omar.Tibolla@astro.uni-wuerzburg.de}

\begin{abstract}
MAGIC (Major Atmospheric Gamma$-$ray Imaging Cherenkov Telescope) is a system of two 17 meters Cherenkov telescopes, sensitive to very high energy (VHE; $> 10^{11}$ eV) gamma radiation above an energy threshold of 50 GeV. The first telescope was built in 2004 and operated for five years in stand‐alone mode. A second MAGIC telescope (MAGIC$-$II), at a distance of 85 meters from the first one, started taking data in July 2009. Together they integrate the MAGIC stereoscopic system. Stereoscopic observations have improved the MAGIC sensitivity and its performance in terms of spectral and angular resolution, especially at low energies. 

We report on the status of the telescope system and highlight selected recent results from observations of galactic and extragalactic gamma$-$ray sources. The variety of sources discussed includes pulsars, galactic binary systems, clusters of galaxies, radio galaxies, quasars, BL Lacertae objects and more.
\end{abstract}


\bodymatter

\section{MAGIC}\label{magic}

\begin{figure} 
\begin{center}
\psfig{file=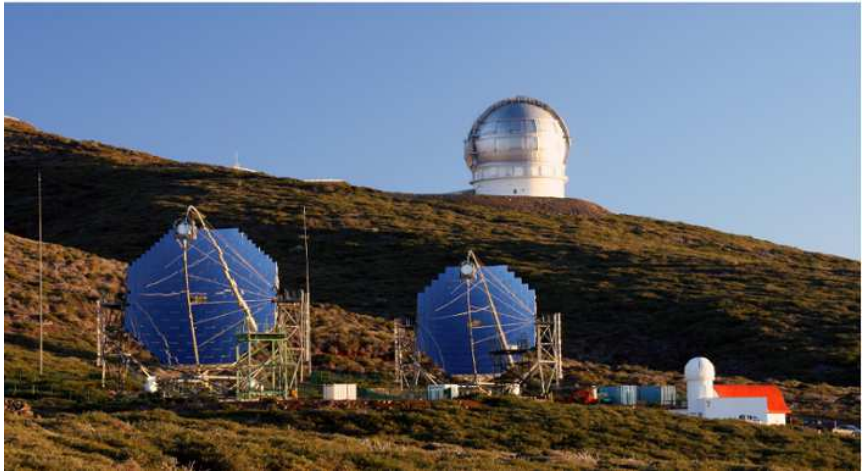,width=4.5in}
\end{center}
\caption{MAGIC Telescopes in Observatorio del Roque de los Muchachos.}
\label{m}
\end{figure}

The Major Atmospheric Gamma$-$ray Imaging Cherenkov Telescope (MAGIC) is a system of two 17$-$meters Atmospheric Cherenkov Telescopes (shown in Fig. \ref{m}) located at the \emph{Observatorio del Roque de los Muchachos} in the island of \emph{La Palma}, 2200 meters above sea level. MAGIC-I has been in operation since 2004 and the stereoscopic system has been operation since 2009. MAGIC has an enhanced duty cycle up to $\sim 17$\% as it is able to operate in presence of moderate moonlight and twilight.

The performances of the MAGIC stereoscopic system are reported in \cite{magic} . The low energy threshold of 50 GeV (or 25 GeV with special trigger setup \cite{sumtrigger} ) allows observations of the distant universe
and overlaps with the Fermi satellite; the angular resolution is $\sim 0.1^{\circ}$ at 100 GeV, down to $\sim 0.05^{\circ}$ above 1 TeV; the energy resolution is 20\% at 100 GeV and goes down to 15\% at 1 TeV.

Another very important feature of MAGIC telescopes is their light structure (ultralight carbon fiber frame), that allows fast repositioning (less than 20 seconds for a 180$^\circ$ repositioning), for fast follow-up observation of gamma-ray bursts (GRBs).

In order to achieve an easier maintenance, better sensitivity and performances (in particular for extended sources), on June 15$^{th}$, the MAGIC telescopes were shut down to perform a major upgrade of the hardware:
\begin{itemize}
 \item {Both telescopes will be equipped with a new 2 GSamples/s readout based on DRS4 chip (linear, low dead time, low noise);}
 \item {The camera of MAGIC-I will be upgraded to a clone of the MAGIC-II camera, i.e. from 577 to 1039 pixels to match the camera geometry and the trigger area of MAGIC-I (currently this is planned for 2012);}
 \item {Both telescopes will be equipped with \emph{sumtrigger} (threshold $\sim$25 GeV \cite{sumtrigger} ) covering the total conventional trigger area (planned for 2012 as well).}
\end{itemize}

The MAGIC scientific program covers different aspects of high energy astrophysics:

\begin{itemize}
 \item {Galactic Objects: Supernovae Remnants (SNRs), Pulsars and Pulsar Wind Nebulae (PWNe).}
 \item {Extragalactic Objects: Active Galactic Nulcei (AGNs), starburst galaxies, clusters of galaxies and Gamma-Ray Bursts (GRBs).}
 \item {Fundamental physics, such as the origin of Cosmic Rays (CRs; that can of course be studied indirectly, by means of studying SNRs for instance, but also directly, considering the showers initiated by primary CRs), Dark Matter (DM) searches and the possible tests of Lorentz invariance violations.}
\end{itemize}

\subsection{Galactic observations}

Recent MAGIC results on Galactic science are here highlighted in four sections:

\begin{itemize}
 \item {Crab Pulsar Wind Nebula.}
 \item {Crab pulsar.}
 \item {Extended sources (PWNe and SNRs).}
 \item {X-ray binaries (XRBs) and Galactic microquasars.}
\end{itemize}

\subsubsection{Crab nebula}

\begin{figure}
\begin{center}
\psfig{file=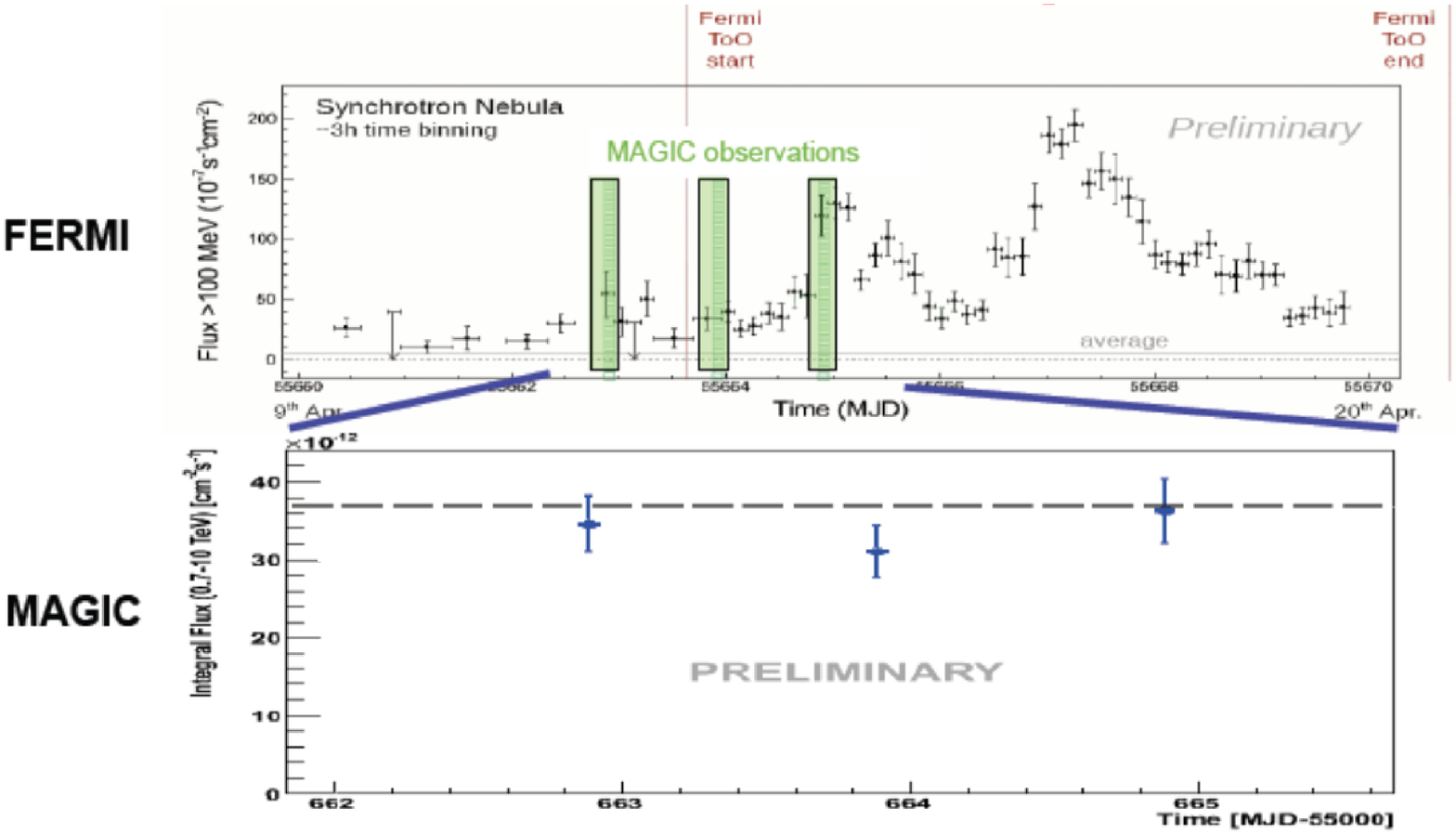,width=4.5in}
\end{center}
\caption{Crab nebula was observed for 3 nights of data in April 2011: during the third night Fermi LAT was observing a flux above 100 MeV that was 15 times higher than the steady flux \cite{crab_icrc}.}
\label{fig_crab}
\end{figure} 

The Crab nebula  is the the prototype of young PWN and it has been considered the ''standard candle`` of Imaging Atmospheric Cherenkov Telescopes (IACTs) so far; however, in the past year, questions have been raised about its flux constancy, after it was seen flaring in GeV gamma-rays (with both \emph{AGILE} and \emph{Fermi LAT}, \cite{111} ), a year-scale longer variability has been observed in X-rays and ARGO-YBJ  \cite{222} reported an increase in the TeV gamma-rays flux. 
MAGIC observed the Crab during the September 2010 flare, finding no indication for variability above 300 GeV \cite{flare} , and during the April 2011 flare (shown in Fig. \ref{fig_crab} and reported in \cite{crab_icrc} ) and no variability observed in the energy range 700 GeV - 10 TeV. However, given the daily binning of MAGIC data, any shorter term variability cannot be excluded so far.

\subsubsection{Crab pulsar}

Most models for gamma-ray emission from pulsars (such as polar cap, outer or slot gap) predict exponential or super-exponential cut-offs in pulsar spectral energy distribution at a few GeV and this is indeed what \emph{Fermi LAT} has observed in the 100 MeV - 10 GeV energy range (e.g. \cite{psr} ) for many pulsars.
Thanks to its low energy threshold MAGIC has the capabilities to test this trend; the Crab pulsar has been observed for 59 hours (between October 2007 and February 2009) with MAGIC-I, allowing the extraction of detailed phase$-$resolved spectra between 25 GeV and 100 GeV: the spectra show a power$-$law behavior and the cut-off extrapolation is ruled out at more than 5 standard deviations \cite{taka} .

After fall 2009, the Crab pulsar was observed for 73 hours in stereoscopic mode; the phase$-$resolved spectra extracted from those observations agree with the ones obtained with MAGIC-I, their simple power law behavior is confirmed and they extend well beyond a cut-off at few GeV energies \cite{crab2} . Moreover MAGIC data are in good agreement with \emph{Fermi LAT} and VERITAS \cite{ver} observations; hence ''standard`` models above mentioned cannot explain this observed behaviour.
Is the Crab pulsar atypical or do other pulsars also have such a VHE power law tail?

\subsubsection{Extended sources}

Thanks to the improved stereoscopic system, MAGIC is more performant also for studying extended sources.

\begin{itemize}
 \item {HESS J1857+026, a VHE unidentified gamma-ray source discovered by H.E.S.S. in 2008 and after suggested to be a PWN powered by the energetic pulsar PSR J1856+0245, has been detected by MAGIC in 2010, allowing us to investigate its energy dependent morphology \cite{1857} . Its spectrum fits well with a power law, consistent with the H.E.S.S. one and its extrapolation; in order to agree with the LAT data a spectral turnover is needed at  10-100 GeV: this could be naturally explained with an Inverse Compton turnover, which would confirm the leptonic nature suggested for this source.}

\item {W51 was detected in GeV gamma-rays energies by \emph{Fermi LAT} \cite{w51f} and at TeV energies by H.E.S.S. \cite{w51h} ; its gamma-ray emission is thought to have its origin in the SNR/MC interaction. MAGIC clearly detected W51 in 2010 (more than 8 standard deviations in 31 hours of observation), confirming its extent ($\sim 0.16^{\circ}$) and the fact that the VHE emission spatially coincides with shocked MC \cite{w51} . Its spectral shape would confirm the hadronic origin of the gamma-ray signal as well.}
\end{itemize}

\subsubsection{XRBs and Galactic microquasars}

Two nice examples of binaries have been observed by MAGIC in the last years:

\begin{itemize}
 \item {The high mass X-ray binary system (HXRB) LSI+61 303 consists of a compact object of unknown nature (i.e. either a Neutron Star or a Black Hole) orbiting around a Be star of 13 solar masses, with a period of $\sim$27 days, and it is located at 2 kpc of distance. It was discovered in VHE gamma-rays by MAGIC in 2005 and regularly monitored since then. 
In 2008, LSI+61 303 faded, however in 2009 we managed to detect it during this low VHE state; and more recently, between Autumn 2010 and Spring 2011, the luminosity of this HXRB was back to the level at which it was first detected \cite{61303} ; eventual correlations with the superorbital periodicity ($\sim$4.6 years) observed in radio are currently under investigations.}

\item {HESS J0632+057 was discovered by H.E.S.S. in 2007; it was the first point-like unidentified source seen at VHE energies and it is the first binary discovery triggered by VHE observations. It was detected with MAGIC in 2011 \cite{mon} in coincidence with a high X-ray activity period. Currently this source is monitored with MAGIC, H.E.S.S. and VERITAS and hence is a very nice example of synergy among different IACTs}
\end{itemize}

However there is another class of XRBs, i.e. objects that show an accretion disk around the compact object and jet-like structures orthogonal to it, the so-called Galactic microquasars, monitored by MAGIC and not detected so far: e.g Cygnus X-1 has been observed for more than 100 hours leading to no detection, and, more recently, also upper limits on GRS1915+105 and Scorpius X-1 have been released by our collaboration \cite{X1} \cite{1915} ; the current upper limits put constraints to the gamma-ray luminosity to be a very small fraction of the kinetic luminosty of the jets.

\subsection{Extragalactic observations}

\begin{figure}
\begin{center}
\psfig{file=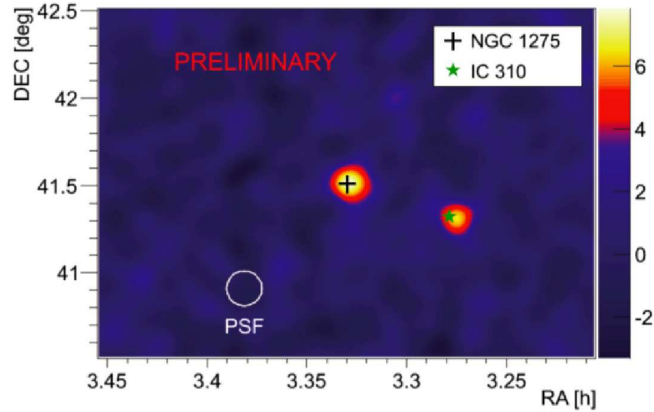,width=3.5in}
\end{center}
\caption{MAGIC preliminary count map of the Perseus cluster region, showing the detection of the two radio galaxies, IC 310 and NGC 1275, in the same field of view \cite{rg}.}
\label{lllll}
\end{figure}

The importance of an improved stereoscopic system reflects also on extragalactic science; in fact in the last 12 months, seven extragalactic objects have been discovered at VHE energies thanks to MAGIC: 4 BL Lac objects (1ES1741+196, 1ES 1215+303, MAGIC J2001+435 and B3 2247+381), 2 radio-galaxies (NGC 1275 and IC 310, visible in Fig. \ref{lllll}) and one Flat Spectrum Radio Quasar (PKS 1221+21).

Another successful strategy in order to discover new extragalactic objects in VHE gamma-rays is represented by the optical trigger, i.e. by monitoring regularly candidate sources by the optical KVA telescope in La Palma (close to MAGIC site) and observing the candidates with MAGIC during their high optical states. This led us to discover several BL Lac objects, such as Mrk180, 1ES1011+496, S5 0716+714, B3 2247+381 and 1ES1215+303.

Another crucial improvement is represented by the complementarity with gamma-ray satellites, that allow us generate much more detailed Spectral Energy Distributions of the sources, by covering in details the High Energy component over 5 decades, and, by means of monitoring these sources also at Radio, optical and X-ray wavelengths, allow us to cover simultaneously more than 17 decades in energy (e.g. \cite{421} ).

However, present day the extragalactic VHE science is no longer restricted to BL Lac objects: MAGIC detected successfully also Radio galaxies (such as M87,  IC 310 and NGC 1275, e.g. \cite{rg} ) and FSRQs \cite{279} \cite{1222} .

\subsubsection{Quasars, the most distant objects}

After the surprising detection by MAGIC, 3C 279 (i.e. the most distant object ever detected at VHE; z=0.536) has been re-observed by MAGIC \cite{279} , confirming that its emission is harder than expected, and showing that the Universe is more transparent to gamma-rays than previously predicted.

This low upper limit to the Extragalactic Background Light (EBL) has been confirmed with the discovery at VHE of another FSRQ: PKS 1222+21 \cite{1222} . Observations of PKS 1222+21 and 3C 279 show the same features:

\begin{itemize}
 \item {Emission up to hundreds of GeV.}
 \item {Fast variability (e.g in PKS 1222+21 we saw 9 minutes doubling times).}
 \item {No signs of intrinsic cut-off.}
\end{itemize}

To reconcile those hard spectra with such a high variability is still a challenge for the theoretical models of photon emission in this type of sources. In fact, in the standard picture \cite{der} \cite{ghi} \cite{sik} , if gamma-rays are produced outside the  Broad Line Region (BLR) by Inverse Compton scattering of dusty torus photons, we can explain explain the smaller-than-expected but it is hard to explain the fast variability; in contrast, if gamma-rays are produced inside the BLR by Inverse Compton scattering of BLR photons, we can explain the variability, but we expect strong absorption and Klein-Nishina suppression (i.e. a cut-off at energies lower than 100 GeV).
More complex models are currently under evaluation, such as strong recollimations of the jet, or the presence of blobs or minijets inside the jet, or the so-called two-zone model (i.e. a large emission zone inside the BLR and in addition a small blob outside).

\subsubsection{GRBs}

As mentioned in section \ref{magic} MAGIC was especially designed to search for prompt emission of GRBs, thanks to its fast repositioning (less than 20 seconds for 180$^{\circ}$) automatically after an alert.
We are observing in average $\sim$1 GRB/month and so far none has been detected at VHE energies. 

Recently, following a X-ray detection, GRB110328 was observed was observed. After a multiwavelength follow-up, it turned out to be hardly classified as a GRB due to its long-lasting activity. The nature of this source is still uncertain \cite{grb} .

\subsection{Fundamental physics}

\subsubsection{Cosmic Rays}

\begin{figure}
\begin{center}
\psfig{file=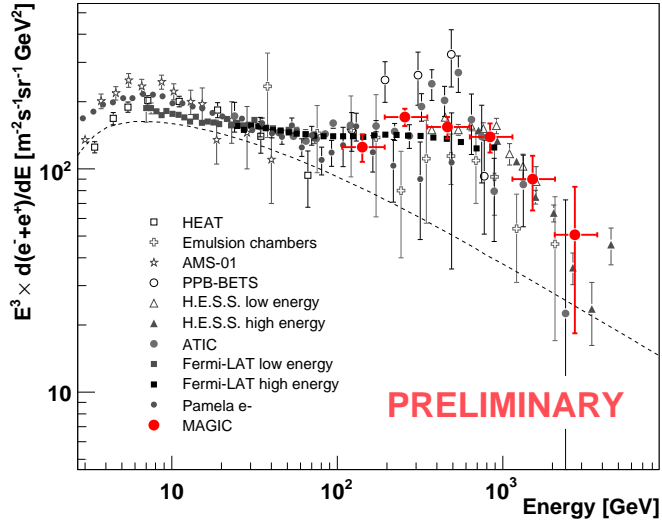,width=3.8in}
\end{center}
\caption{The e$^{\pm}$ CR spectrum measured by MAGIC is in good agreement with the previous measurements \cite{e} .}
\label{ele}
\end{figure}

The first MAGIC results on CR electrons spectrum, visible in Fig. \ref{ele}, are based on 14 hours of data taken in 2009-2010: the e$^{\pm}$ spectrum has been measured in the energy range between 100 GeV and 3 TeV.
The energy distribution agrees well with the previous measurements of H.E.S.S. and Fermi LAT (and the peak detectedby ATIC cannot be excluded or confirmed) \cite{e} .

A related initiative (following the experience of the ARTEMIS experiment) consists in probing the e$^{+}/$e$^{-}$ ratio at 300-700 GeV by measuring the shadowing of the CR flux by the Moon.
This measurable effect (for 50 hours of observations) has been estimated to be $\sim$4.4\% of the Crab flux in the range 300-700 GeV; given its small observability window (i.e. the most favorable observation periods are the spring equinox and the autumn equinox for e$^{+}$ and e$^{-}$ respectively) this measure could be possible with MAGIC integrating data over a few years (for details \cite{e2} ).

\subsubsection{Dark Matter: indirect searches} Super-Symmetrical (SUSY) extensions of the standard model foresee the existence of stable, weakly interacting particles (e.g. the lightest neutralino) which could account for part of the dark matter in the Universe. The annihilation of neutralinos may give rise to gamma-rays in the energy range accesible to MAGIC.
Such signals have been sought with MAGIC in several targets:

\begin{itemize}
 \item {Galaxy Clusters. MAGIC searches concentrated on the Perseus cluster, that is really challenging for several reasons: (1) for the presence of NGC 1275 and IC 310, (2) since the expected flux much smaller than the one coming from CRs and moreover (3) it could possibly have a very extended DM profile.}
 \item {Unidentified Fermi Objects. 1FGL J0338.8+1313 and 1FGL J2347.3+0710 show a hard spectrum in \emph{Fermi LAT} data and they could be DM micro-spikes in the Galactic halo; they have been observed for relatively short exposures ($\sim$10 hours) and lead to no detection so far \cite{dm} .}
 \item {Dwarf spheroidal galaxies. Several of them were observed in the past (such as Draco and Willman-1) and recently Segue-1, i.e. the most DM dominated object known so far ($M/L > 1000)$, has been observed for 29 hours, showing no significant excess \cite{segue} .}
\end{itemize}

DM indirect searches leaded to no detections so far and the derived upper limits are still above theoretical expectations.

\section{Acknowledgments}
The MAGIC Collaboration would like to thank the Instituto de Astrofisica de Canarias for the excellent working conditions at the Observatorio del Roque de los Muchachos in La Palma. The support
of the German BMBF and MPG, the Italian INFN, the Swiss National Fund SNF, and the Spanish MICINN is gratefully acknowledged. This work was also supported by the Marie Curie program, by the CPAN CSD2007-00042 and MultiDark CSD2009-00064 projects of the Spanish Consolider-Ingenio 2010 programme, by grant DO02-353 of the Bulgarian NSF, by the YIP of the Helmholtz Gemeinschaft, by the DFG Cluster of Excellence ''Origin and Structure of the Universe``, by the DFG Collaborative Research Centers SFB823/C4 and SFB876/C3, and by the Polish MNiSzW grant 745/N-HESS-MAGIC/2010/0.

\end{document}